\documentclass[pra,11pt,tightenlines,superscriptaddress]{revtex4}
\usepackage{amsmath,amssymb,amsfonts,graphicx}
\usepackage[colorlinks=true, linkcolor=blue, citecolor=magenta]{hyperref}
\usepackage{color}

\def\>{\rangle}
\def\<{\langle}
\def\({\left(}
\def\){\right)}

\newcommand{\eref}[1]{Eq.~\eqref{#1}}

\def\reals{\mathbb{R}}

\def\th{^{\rm th}}
\def\cO{\mathcal{O}}

\renewcommand{\vec}[1]{\mathbf{#1}}
\newcommand{\mat}[1]{\mathsf{#1}}
\def\id{\mat 1} 

\begin{document}


\title{The Optical Frequency Comb as a One-Way Quantum Computer}

\author{Steven T.\ Flammia}
\email{sflammia@perimeterinstitute.ca}
\affiliation{Perimeter Institute for Theoretical Physics, Waterloo, Ontario N2L 2Y5, Canada}

\author{Nicolas C.\ Menicucci}
\email{nmenicucci@perimeterinstitute.ca}
\affiliation{Perimeter Institute for Theoretical Physics, Waterloo, Ontario N2L 2Y5, Canada}
\affiliation{Department of Physics, Princeton University, Princeton, New Jersey 08544, USA}
\affiliation{Department of Physics, University of Queensland, Brisbane, Queensland 4072, Australia}

\author{Oliver Pfister}
\email{opfister@virginia.edu}
\affiliation{Department of Physics, University of Virginia, Charlottesville, Virginia 22903, USA}

\date{November 15, 2008}

\begin{abstract}
In the one-way model of quantum computing, quantum algorithms are implemented using only measurements on an entangled initial state.  Much of the hard work is done up-front when creating this universal resource, known as a cluster state, on which the measurements are made.  Here we detail a new proposal for a scalable method of creating cluster states using only a single multimode optical parametric oscillator (OPO).  The method generates a continuous-variable cluster state that is universal for quantum computation and encoded in the quadratures of the optical frequency comb of the OPO.  This work expands on the presentation in Phys.\ Rev.\ Lett.~\textbf{101}, 130501~(2008).

\end{abstract}

\maketitle

\section{Introduction}

The invention of one-way quantum computing (QC) introduced a new paradigm for quantum information processing~\cite{Raussendorf2001}, whereby the unitary evolution required in the traditional circuit model~\cite{Nielsen2000} is replaced by measurements on a highly entangled {\em cluster state\/}~\cite{Briegel2001}.  In this model, quantum information exists virtually in the cluster state and is manipulated through adaptive single-qubit projective measurements.  The measurement backreaction on the entangled state effectively induces dynamics, and hence quantum computation.  In this way, the cluster state acts as a quantum field-programmable gate array, or a quantum ``substrate''~\cite{Raussendorf2001}, on which a quantum circuit is inscribed and simultaneously evaluated via measurements.

Because it eliminates the issue of dynamic control, one-way QC is an especially attractive model for certain physical systems where creating cluster states is natural.  Natural, in this context, means that large cluster states (say, with more than 100 qubits) can be created in just a single step, without the need to build the cluster piecewise before the computation begins.  The most prominent example of such a system is an optical lattice with cold atoms~\cite{Brennen1999, Mandel2003}, but other proposals exist, e.g.,\ using solid-state qubits~\cite{You2007}.  These systems provide ambitious ``top-down'' approaches to quantum computing, where the issue of scaling the number of qubits is addressed up front.  This is in contrast to the more widely attempted ``bottom-up'' approaches which initially emphasize one- and two-qubit control~\cite{Nielsen2004,Browne2005,Barrett2005}, such as ion traps~\cite{Cirac1995}, linear optics~\cite{Knill2001}, and solid state implementations \cite{Kane1998, Loss1998}, among others~\cite{Roadmap}.

In this paper, we provide a detailed proposal for another candidate for creating large cluster states efficiently: a single optical parametric oscillator~(OPO) driven by a multifrequency pump.  This result was first reported in Ref.~\cite{Menicucci2008}.  This article is an exhaustive description that fills in details of the derivations and proposed implementations.

The eigenmodes of an optical cavity (e.g.\ two mirrors facing each other), whose frequencies satisfy a constructive interference requirement, define an \emph{optical frequency comb}~(OFC), so called because the frequencies are evenly spaced.  The OFC has outstanding classical coherence properties, as illustrated by the revolutionary and now ubiquitous use of million-frequency, mode- and carrier-envelope-phase-locked femtosecond lasers in time and frequency metrology~\cite{Hansch2006,Hall2006}. When the electromagnetic field is quantized, this comb of eigenmodes forms a set of independent harmonic oscillators which we propose to use as the core of a quantum computer by entangling them into a cluster state.  The goal being to generate cluster-state entanglement between these quantum modes, or \emph{qumodes}~\footnote{The term `qumode' was suggested to us by Daniel Gottesman and is used, in analogy with `qubit', to emphasize that the modes we're working with are quantum in nature---that is, they have noncommuting raising and lowering operators associated to them instead of classical amplitudes and phases.}, we first augment the optical cavity with a nondissipative, second-order nonlinear gain medium, thus converting it into an optical parametric oscillator~(OPO).  With an appropriate nonlinear medium and a pump beam with the right frequency content, the comb of qumodes will encode a large square lattice continuous-variable cluster state~\cite{Menicucci2008}.  Remarkably, this can be achieved using just a single OPO---and with a \emph{constant} number of pump frequencies, regardless of the size of the cluster.  Compared to existing proposals~\cite{Menicucci2006,vanLoock2007}, these modest experimental requirements, together with the naturally large set of qumodes inside the comb mean that this approach has strong potential for scalability.  Detailing this construction is our main goal.

The paper is organized as follows.  In Section~\ref{S:OPOs}, we discuss the basic physics of the OPO.  In Section~\ref{S:CVCS} we define continuous-variable (CV) cluster states, drawing distinctions with their qubit counterparts, and discuss previous literature dealing with the creation of CV cluster states.  In Section~\ref{S:CVCSfrom1OPO} we outline the program of creating a CV cluster state from just a single OPO and a multifrequency pump~\cite{Menicucci2007} and introduce a simplifying ansatz~\cite{Zaidi2008}.  We demonstrate the power of this approach in Section~\ref{S:ring}, where we detail the construction of a ring-shaped CV cluster state that is universal for single-mode transformations.  The ring construction contains all of the essential ideas for the main result in Section~\ref{S:torus}, which was first discussed in Ref.~\cite{Menicucci2008}: a fully universal CV cluster state from a single OPO using only 15~pump frequencies, regardless of the size of the cluster.  We detail a proposed experimental implementation of this scheme in Section~\ref{S:exp}.  The important but unresolved issues of error correction and fault tolerance are discussed in Section~\ref{S:qecft}, and we conclude in Section~\ref{S:conclusion}.

\section{Optical Parametric Oscillators}\label{S:OPOs}

When an OPO is driven by a pump field with frequency and wave vector $(\omega_p,\vec k_p)$, the nonlinear medium inside promotes the up- or downconversion of pump photons into pairs of signal photons [($\omega_m,\vec k_m$);($\omega_n,\vec k_n$)], if the phasematching conditions
\begin{align}
\label{E:photonEC}
\left\{\begin{array}{c@{=}c}
	\omega_p&\omega_m+\omega_n\\
	\vec k_p&\vec k_m+\vec k_n \end{array}\right.
\end{align}
hold, in particular so long as the photons have frequencies within the phasematching bandwidth of the medium. This interaction is also dependent on other optical quantum numbers, such as the polarization and the transverse spatial mode. We assume for now that only one polarization per signal frequency gets phasematched for a given pump frequency; we will relax this restriction later. The phasematching is assumed collinear.

It is by now well established that such interactions lead to bipartite CV entanglement between the signal qumodes~\cite{Braunstein2005a} both below~\cite{Ou1992a} and above~\cite{Villar2005,Su2006,Jing2006} the OPO threshold.  Any single such conversion, a.k.a.\ parametric, process has an interaction-picture Hamiltonian of the form~$\mathcal{H} = i \hbar \kappa (a^\dag_m a^\dag_n - a^{\phantom\dag}_m a^{\phantom\dag}_n)$, where $\kappa=\beta\chi$ is a nonlinear coupling strength (squeezing parameter per unit time), with $\chi$ the nonlinear susceptibility of the medium and $\beta$ the complex amplitude of the pump (assumed undepleted and classical), and $a^\dag_n$ is the creation operator for photons with frequency~$\omega_n$. We define the amplitude and phase quadrature operators for the $k\th$ qumode as $q_k= (\hat a^\dag_k+\hat a^{\phantom \dag}_k)/\sqrt{2}$ and $p_k = i(\hat a_k^\dag-\hat a^{\phantom \dag}_k)/\sqrt{2}$. The sign of $\kappa$ is set by the pump phase equal to 0, in which case the Hamiltonian $\mathcal{H}$ generates quantum squeezing of the phase-sum and amplitude-difference simultaneously, which constitutes a realization of the Einstein-Podolsky-Rosen paradox \cite{Einstein1935,Reid1988} and therefore bipartite entanglement. Similarly, if the pump phase is $\pi$, there will be simultaneous amplitude-sum and phase-difference squeezing.  Note also that these two possibilities coincide respectively with parametric amplification and deamplification of the signal fields when they are seeded by coherent states.

A single pump frequency can simultaneously generate multiple pairwise interactions, and therefore multipartite entanglement, if there are multiple pairs of signal frequencies that satisfy Eq.~\eqref{E:photonEC} within the phasematching bandwidth of the crystal~\cite{Pfister2004,Bradley2005,Zaidi2008}.  In this case, we can write the full interaction-picture Hamiltonian as
\begin{align}
\label{eq:Hamiltonian}
	\mathcal{H}(\mat G) = i \hbar \kappa \sum_{m,n} G_{mn} (a^\dag_m a^\dag_n - a^{\phantom\dag}_m a^{\phantom\dag}_n)\;.
\end{align}
This Hamiltonian is written explicitly as a function of a matrix $\mat G$ (with entries~$G_{mn}$), which is the adjacency matrix describing the pairwise coupling strength between each pair of signal modes in units of $\kappa$.  The matrix $\mat G$ is real and symmetric, with positive (negative) entries corresponding to down- (up-)conversion.

So far, this is a fairly general Hamiltonian, since we have said nothing about the ordering of the signal modes.  In an OPO the signal modes are equally spaced in frequency, forming what's called an \emph{optical frequency comb}, so we can choose to order them sequentially like this:
\begin{align}
	\omega_n = \Omega + n \Delta \omega\;,
\end{align}
where $\Omega$~is an offset frequency (discussed below), and $\Delta \omega$~is the free spectral range of the OPO (i.e.,\ the mode spacing). We neglect the effects of optical dispersion in this paper.  Not all values of~$n$ or~$m$ correspond to entries within $\mat G$, however, since the only modes involved in the interaction are those within the phasematching bandwidth.  We therefore choose~$\Omega$ such that $\omega_1$ corresponds to the lowest-frequency mode within the phasematching bandwidth and $\omega_N$~to the highest, which allows us to label the rows and columns of~$\mat G$ each from~1 to~$N$ (inclusive), corresponding exactly to the cavity modes available for interaction.  With this convention, the phasematching condition of Eq.~\eqref{E:photonEC} becomes~$n + m = \text{(const.)}$, which selects a skew-diagonal band to be the only nonzero entries in $\mat G$.  (The particular band selected is determined by the relationship of the pump frequency to the phasematching bandwidth.)  The entries in this band are all equal and given in units of~$\kappa$, with the sign indicating downconversion if positive or upconversion if negative, as above.

\begin{figure}
\begin{center}
\includegraphics[width=\columnwidth]{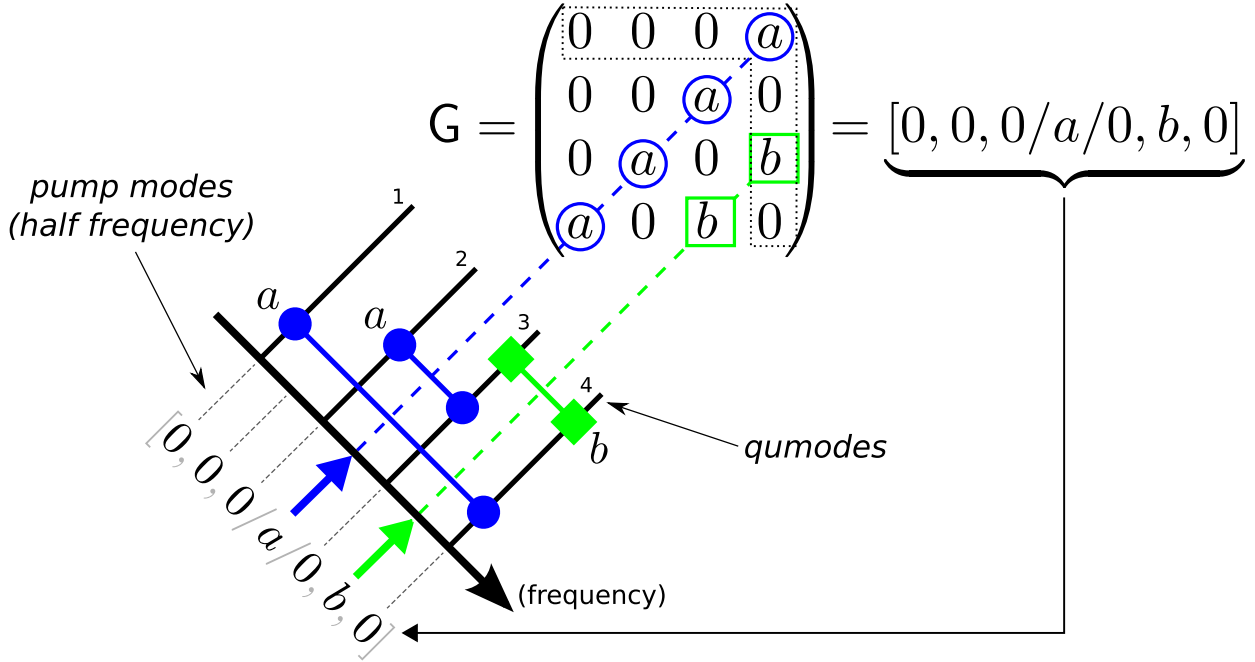}
\caption{Hankel shorthand and pump specification.  A Hankel matrix is uniquely specified by the entries along the top and down the right side.  We collect these entries into a shorthand vector, representing the entire matrix itself, with the top-right entry set off with slashes.  When $\mat G$ is Hankel, its shorthand vector immediately specifies the pump spectrum required to implement $\mathcal H(\mat G)$.  Each nonzero entry in the shorthand vector denotes the amplitude of a frequency in the pump, each of which generates CV entanglement between pairs of qumodes in the OFC symmetric about half that frequency.  This accounts for all couplings prescribed by~$\mat G$.}
\label{F:Hankel}
\end{center}
\end{figure}

This construction generalizes easily to the case where the OPO is pumped with multiple frequencies simultaneously.  In this case, each pump frequency corresponds to a different phasematching condition, Eq.~\eqref{E:photonEC}.  Thus, each corresponds to a different skew-diagonal band in~$\mat G$.  Since we want all such interactions summed together as in Eq.~\eqref{eq:Hamiltonian}, this leads us to consider exclusively those $\mat G$'s that are of \emph{Hankel form}---i.e.\ those that have constant skew-diagonals.  For convenience later, we introduce a shorthand notation for Hankel matrices of any size.  A Hankel matrix is completely specified by its first row and last column, so we will collect these terms into a vector, setting aside by slashes the middle element common to both.  The shorthand form of a given Hankel matrix is unique and represents the entire matrix itself.  As such, the following is a literal equality:
\begin{align}
\label{eq:shorthandexample}
	[a,b/c/d,e] =
	\begin{pmatrix}
		a &	b &	c \\
		b &	c &	d \\
		c &	d &	e
	\end{pmatrix}
	\;.
\end{align}
While this shorthand form can be used to represent any Hankel matrix, it has an additional advantage when used to represent~$\mat G$ in that it corresponds simply and directly to the spectrum of the generating pump beam, as illustrated in Figure~\ref{F:Hankel}.

\subsection*{Polarization and Block-Hankel Matrices}

The description above assumes that only one interaction per pump frequency is phasematched in the crystal.  Appropriate additional quantum numbers (polarization, spatial mode, etc.)\ for the pump and signal modes are implied.  An important extension of this scheme is when there are multiple modes that are frequency degenerate and yet phasematched independently by the crystal.  The simplest example of this is when the crystal phasematches multiple polarization-dependent interactions, but other possibilities include wavevectors and transverse spatial modes.  The simultaneous quasiphasematching \cite{Armstrong1962,Fejer1992} of four polarization-sensitive interactions ZZZ, ZYY, and YZY/YYZ (where the first letter is pump polarization) has been demonstrated experimentally in a periodically poled $\mathrm{KTiOPO_4}$ (PPKTP) crystal~\cite{Pooser2005}.

In such a case, the Hankel restriction on $\mat G$ can be relaxed to \emph{block-Hankel}, with the block size given by the number of frequency degenerate modes.  Take the example of polarization-dependent interactions discussed above.  If $M$ frequencies are phasematched by the crystal, with two polarization-modes each, then there are $N = 2M$ total modes, ordered $1_\text{Y}, 1_\text{Z}, 2_\text{Y}, 2_\text{Z}, \dotsc, M_\text{Y}, M_\text{Z}$, with~`Y' and~`Z' labeling the two polarizations.  Extensions including more quantum numbers are possible for setups that involve more complicated frequency-degenerate phasematching.  The important thing is that each frequency label is completed by a set of additional mode labels, with the same ordering of additional quantum numbers within each set.

\begin{figure}
\begin{center}
\includegraphics[width=.75\columnwidth]{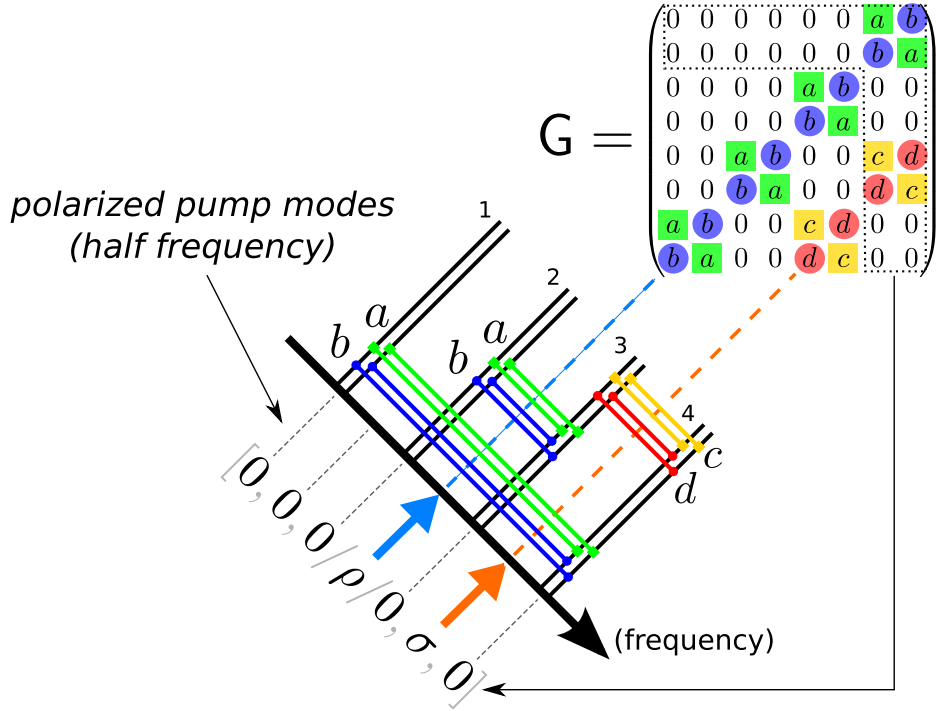}
\caption{Polarization-sensitive interactions and block-Hankel matrices.  This is a generalization of Figure~\ref{F:Hankel} to include polarization-sensitive interactions.  Each of the four frequencies now has two independent polarization-modes that can be coupled within the crystal simultaneously.  The Hamiltonian adjacency matrix~$\mat G$ is now $8 \times 8$ and is no longer strictly of Hankel form but rather is block-Hankel with $2 \times 2$~blocks.  These blocks along the top and right side can still be collected into a shorthand vector, as shown, where~$\rho = \left(\protect\begin{smallmatrix}a & b \\ b & a\protect\end{smallmatrix}\right)$, $\sigma = \left(\protect\begin{smallmatrix}c & d \\ d & c\protect\end{smallmatrix}\right)$, and~0 represents a block of zeros.  The nonzero blocks each specify the amplitude and polarization angle in the (YZ)-plane of a single pump frequency when used with a PPKTP crystal~\cite{Pooser2005}.}
\label{F:blockHankel}
\end{center}
\end{figure}

Returning to the simplest case of polarization-dependent phasematching (which is all we will use in this paper), a single pump frequency can now generate several interactions at once, depending on its polarization.  Using the crystal discussed above~\cite{Pooser2005}, a monochromatic pump implements a skew-diagonal band of $2 \times 2$ blocks in~$\mat G$.  The particular form of the block is determined by the pump's polarization angle in the (YZ)-plane.  Multiple polarized frequencies (possibly at different angles) in a single pump will produce a block-Hankel~$\mat G$, again with $2 \times 2$ blocks, as illustrated in Figure~\ref{F:blockHankel}.  Furthermore, the Hankel shorthand discussed above may be extended intuitively to block-Hankel matrices.  The shorthand vector is now comprised of blocks instead of single entries and, when applied to $\mat G$, still directly specifies the pump spectrum, including the necessary polarization of each frequency in the beam.

\section{Continuous-Variable Cluster States}\label{S:CVCS}

The relationship between the qubit-based and CV-based formalisms is by now well-established, and spans the areas of QC~\cite{Lloyd1999,Bartlett2002}, quantum error correction~\cite{Braunstein1998a,Gottesman2001}, cluster states~\cite{Zhang2006}, and one-way QC~\cite{Menicucci2006}.  Given a collection of $N$ qumodes, we collect the single-mode quadrature operators into column vectors given by $\vec q = (q_1, \dotsc, q_N)^T$ and $\vec p = (p_1, \dotsc, p_N)^T$.  Then the definition of a CV cluster state~\cite{Zhang2006,vanLoock2007a,Menicucci2006} is any Gaussian state whose quadratures satisfy, in the limit of large squeezing,
\begin{align}
\label{eq:CVCSdef}
	\vec p(r) - \mat A\,\vec q(r) \xrightarrow{r\to\infty} \vec 0\;,
\end{align}
where $r>0$ is a squeezing parameter and $\mat A$ is the adjacency matrix of the CV cluster's graph.  Because they are equivalent descriptions, we can speak interchangeably about the graph of the cluster state and its adjacency matrix $\mat A$.  The usual choice of $\mat A$ for universal cluster-state QC is a square-lattice graph, but others are possible as well.  Many regular lattices~\cite{Van-den-Nest2006} and also irregular lattices above a percolation threshold~\cite{Kieling2007, Browne2008}, and even certain more exotic states~\cite{Gross2007} are now known to be universal resource states.

The infinite-squeezing limit of \eref{eq:CVCSdef} is an important difference from the standard (discrete-variable) cluster state; it is not achievable by any finite-energy state, so we must be content with finitely (albeit strongly) squeezed states as an approximation.  This means that errors due to finite squeezing are inherent in CV cluster state QC, a point we return to in Section~\ref{S:qecft}.

Previous proposals for creating CV cluster states called for $\cO(N)$ vacuum squeezers and  $\cO(N)$ quantum non-demolition gates, which can be implemented by experimentally arduous inline squeezers (e.g.,\ seeded OPOs)~\cite{Yurke1985,Braunstein2005}, although realization with more convenient offline, vacuum squeezers was recently achieved \cite{Yoshikawa2008}. An elegant simplification is provided by the Bloch-Messiah decomposition \cite{Braunstein2005,vanLoock2007}, which reduces the experimental requirements to $N$ OPOs followed by a $2N$-port interferometer.  This method has already been used successfully to demonstrate the creation and use of four-mode CV cluster states for simple quantum information processing tasks~\cite{Yonezawa2008,Yukawa2008,Su2007}.  In contrast, our method requires a \emph{single~OPO} and \emph{no~interferometer}. As we shall see, the pumping field's complexity is a constant with~$N$. The required nonlinear medium is sophisticated yet feasible with current technology and has been demonstrated experimentally~\cite{Pooser2005}.

Quantum computation on the CV cluster state~\cite{Menicucci2006} proceeds analogously to the case of qubit cluster states~\cite{Raussendorf2001}.  Adaptive measurements are performed on the cluster to induce computation.  Homodyne detection alone is sufficient to implement all Gaussian operations, which are the analogs of Clifford gates for qubits.  In addition, the required basis for a given homodyne measurement does not depend on the outcomes of other homodyne measurements.  Thus, measurements implementing Gaussian operations can be performed in any order, and even in parallel.  The availability of a non-Gaussian measurement (e.g.,\ photon counting) is essential for universal QC~\cite{Menicucci2006}. Non-Gaussian measurements, like non-Clifford gates for qubits,  are not parallelizable in general because future measurement bases will depend on the outcome of this measurement.

\section{CV Cluster States From a Single OPO}\label{S:CVCSfrom1OPO}

In this section, we review the relationship between the coupling matrix $\mat G$ of the OPO Hamiltonian in \eref{eq:Hamiltonian}, and the adjacency matrix $\mat A$ of the created CV cluster state in \eref{eq:CVCSdef}.  Our treatment differs from the one in Ref.~\cite{Menicucci2007} in that we make an additional simplifying ansatz that was first made explicit in Ref.~\cite{Zaidi2008}.  This analysis also makes explicit the scaling with respect to the squeezing parameter in \eref{eq:CVCSdef}.  A more general analysis of the underlying mathematical structures can be found in Ref.~\cite{Flammia2008}. 

As we have seen in Section~\ref{S:OPOs} based on physical arguments, we should specialize to $\mat G$'s that are Hankel or block-Hankel, as in Figures~\ref{F:Hankel} and~\ref{F:blockHankel}.  We make the simplifying ansatz that there are no degenerate squeezing interactions within the OPO, which, with the (block-)Hankel requirement ensures that $\mat G$ is the adjacency matrix of a \emph{bipartite} graph.  Any bipartite~$\mat G$ can be factored into a tensor product as
\begin{align}
	\mat G = \mat A_0 \otimes \mat X
\end{align}
where $\mat X = \big(\begin{smallmatrix} 0&1\\1&0 \end{smallmatrix}\big)$, and $\mat A_0$ is a block-Hankel matrix having the same size blocks as those of~$\mat G$.

For example, consider the configuration depicted in Fig.~\ref{F:Hankel}.  We can factor $\mat G$ as
\begin{align}
	\mat G = [0,0,0/a/0,b,0] =
	\begin{pmatrix}
		0 & 0 & 0 & a \\
		0 & 0 & a & 0 \\
		0 & a & 0 & b \\
		a & 0 & b & 0 \\
	\end{pmatrix}
	=
	\begin{pmatrix}
		0 & a \\
		a & b 
	\end{pmatrix}
	\otimes
	\begin{pmatrix}
		0 & 1 \\
		1 & 0 
	\end{pmatrix}
	= 
	[0/a/b]\otimes [0/1/0]
	=
	\mat A_0 \otimes \mat X
	\;,
\end{align}
where we move freely between the matrix form of $\mat G$ and the Hankel shorthand discussed above.  A Hankel $\mat G$ with an even number of modes has such a factorization if and only if it contains no single-mode squeezing---or equivalently, if it has only zeros on the main diagonal.


We now introduce the additional ansatz of Ref.~\cite{Zaidi2008} and assume that $\mat G^T \mat G = \mat G^2 = \id$, or equivalently, that $\mat A_0^2 = \id$.  (Here and throughout, the size of the identity matrix represented by the symbol~$\id$ is given by the context.  The same holds for the zero matrix denoted by $0$.)  By assuming that $\mat G$ is orthogonal, we greatly simplify its relationship to $\mat A$.  In fact, we will simply define
\begin{align}\label{E:switch}
	\mat A := \mat X \otimes \mat A_0
\end{align}
and show that it is the adjacency matrix for the CV cluster state generated by $\mathcal H(\mat G)$~\cite{Menicucci2007}.  This definition would imply that $\mat A$ and $\mat G$ are isomorphic graphs, since reversing the tensor product order of an adjacency matrix corresponds merely to a particular permutation of node labels.  Continuing with the example from above,
\begin{align}
	\mat A = \mat X \otimes \mat A_0 = 
	\begin{pmatrix}
		0 & \mat A_0 \\
		\mat A_0 & 0 
	\end{pmatrix}
	= 
	\begin{pmatrix}
		0 & 0 & 0 & a \\
		0 & 0 & a & b \\
		0 & a & 0 & 0 \\
		a & b & 0 & 0 \\
	\end{pmatrix}
	\cong
	\begin{pmatrix}
		0 & 0 & 0 & a \\
		0 & 0 & a & 0 \\
		0 & a & 0 & b \\
		a & 0 & b & 0 \\
	\end{pmatrix}
	=
	\mat A_0 \otimes \mat X = \mat G
	\;.
\end{align}
We denote equivalence up to permutation by $\cong$, which in this case involves exchanging the labels of modes~2 and~3.

To show this correspondence, we begin by collecting the quadrature operators $\vec q$ and $\vec p$ into a single concatenated vector $\vec v = (q_1, \ldots , q_N , p_1 , \ldots , p_N)^T$ which represents the vacuum (i.e.,\ their measurement statistics are those of the vacuum state).  In the Heisenberg picture, symplectic linear transformations on this vector correspond to Gaussian transformations in Hilbert space~\cite{Simon1994}.  The unitary operation $\exp(- i t \mathcal{H})$ of the OPO Hamiltonian from \eref{eq:Hamiltonian} acts on the qudrature operators (in the Heisenberg picture) through the following linear symplectic matrix acting on $\vec v$:
\begin{align}
	\mat U = 
	\begin{pmatrix}
		e^{r \mat G} & 0 \\
		0 & e^{-r \mat G}
	\end{pmatrix}
	\; ,
\end{align}
where $r = \kappa t$.  We additionally allow permutation of the qumode labels and phase-shifting of individual qumodes; these are completely passive operations and don't change the required physical setup used to generate the state.  Using the permutation freedom, we choose to reorder the quadratures so as to reverse the tensor product order, as in \eref{E:switch}, effectively converting $\mat G$ to $\mat A$, and resulting in the redefinition
\begin{align}
	\mat U \to
	\begin{pmatrix}
		e^{r \mat A} & 0 \\
		0 & e^{-r \mat A}
	\end{pmatrix}
	\;.
\end{align}
Using $\mat G^2 = \mat A^2 = \id$, we can evaluate $\mat U$ using the identity $e^{\pm r \mat A} = \cosh(r) \id \pm \sinh(r) \mat A$, 
resulting in
\begin{align}
	\mat U =
	\begin{pmatrix}
		c \id	&		s \mat A_0 &	0 &			0 \\
		s \mat A_0 &	c \id &		0 &			0 \\
		0 &			0 &			c \id	&		-s \mat A_0 \\
		0 &			0 &			-s \mat A_0 &	c \id
	\end{pmatrix}
	\;,
\end{align}
where $c = \cosh(r)$, $s = \sinh(r)$, and each block is the same size as $\mat A_0$.  We now phase shift  half of the qumodes by $-\pi/2$ using the symplectic matrix
\begin{align}
	\mat T = \begin{pmatrix}
		\id & 0 & 0 & 0 \\
		0 & 0 & 0 & -\id \\
		0 & 0 & \id & 0 \\
		0 & \id & 0 & 0
	\end{pmatrix}\;.
\end{align}
This operation does not need to be implemented actively since it is simply a redefinition of the quadratures ($q \to -p$ and~$p \to q$) for the phase-shifted modes.  This results in
\begin{align}
	\mat T \mat U =
	\begin{pmatrix}
		c \id	&		s \mat A_0 &	0 &			0 \\
		0 &			0 &			s \mat A_0 &	-c \id \\
		0 &			0 &			c \id	&		-s \mat A_0 \\
		s \mat A_0 &	c \id &		0 &			0 
	\end{pmatrix}
	\;.
\end{align}
To verify that \eref{eq:CVCSdef} holds, it is sufficient to check that when we act on the transformation matrix~$\mat T \mat U$ with the rectangular matrix
\begin{align}
	\begin{pmatrix}
		-\mat A & \id 
	\end{pmatrix}
	= 
	\begin{pmatrix}
		0 &			-\mat A_0 &	\id &		0 \\
		-\mat A_0 & 	0 &			0 &		\id
	\end{pmatrix}
	\; ,
\end{align}
we get a matrix that vanishes as $r \to \infty$.  We find that, in fact,
\begin{align}
	\begin{pmatrix}
		-\mat A & \id 
	\end{pmatrix} \mat T \mat U = e^{-r}
	\begin{pmatrix}
		0 & 0 & \id & \mat A_0 \\
		-\mat A_0 & \id & 0 & 0
	\end{pmatrix} \xrightarrow{r\to\infty} 0 \; .
\end{align}
This shows that the CV cluster-state relation, \eref{eq:CVCSdef}, is satisfied.

We have shown that if a Hamiltonian coupling matrix $\mat G$ can be found that satisfies the (rather restrictive) assumptions above---namely, that it be (block-)Hankel, orthogonal, and bipartite---then it creates a CV cluster state with an identical adjacency matrix, up to permutation of labels and local redefinition of quadratures.  In the next two sections, we will provide examples of such $\mat G$'s that yield CV cluster states with very interesting graphs using only a constant number of pump frequencies.    In Section \ref{S:ring}, we show how to construct a ring graph which is universal for single mode transformations, and in Section \ref{S:torus} we show how to construct a torus, which is fully universal for quantum computation.  In what follows, since they are equivalent, the distinction between~$\mat G$ and~$\mat A$ will be eliminated and $\mat A$~used to represent both.

\section{Single-Mode-Universal CV Cluster State}\label{S:ring}

We first illustrate the main ideas on a simpler graph that is universal for \emph{single-mode} operations (but not for universal QC, which requires multimode operations).  We desire such a CV cluster state that also has a bicolorable graph~\cite{Menicucci2007} and whose adjacency matrix is orthogonal and Hankel for experimental simplicity~\cite{Zaidi2008}.  Orthogonality of the adjacency matrix ($\mat A \mat A^T = \mat 1$) for an undirected graph ($\mat A = \mat A^T$) yields $\mat A^2 = \mat 1$, or
\begin{align}
\label{eq:orthograph}
	(\mat A^2)_{jk} = \sum_l A_{jl} A_{lk} = \delta_{jk}\;.
\end{align}
Eq.~\eqref{eq:orthograph} has a geometric interpretation: $(\mat A^n)_{jk}$ represents the sum of the weights of all $n$-length paths from node~$j$ to node~$k$, where the weight of such an ``$n$-path'' equals the product of all edge weights along the path.  Eq.~\eqref{eq:orthograph} enforces two conditions:
\begin{enumerate}
\item[(1)] all 2-paths that begin and end on the same node have weights that sum to~1, and
\item[(2)] all 2-paths that link different nodes have weights that cancel out.
\end{enumerate}
The simplest graph to try would be a line graph~\cite{Raussendorf2001} but such a graph necessarily fails condition~(1) because of irregularity at the boundaries. The next natural graph to consider is therefore a ring graph, which is a regular graph, but orthogonality is still prohibited because each node is connected to its next-nearest neighbor by only one 2-path~\footnote{The exception is a four-node ring.  This is unimportant since we want a graph with many nodes.}, for which the sum in Eq.~\eqref{eq:orthograph} collapses to a single term (which must be nonzero), thus violating condition~(2).

\begin{figure}
\begin{center}
\includegraphics[width=\columnwidth]{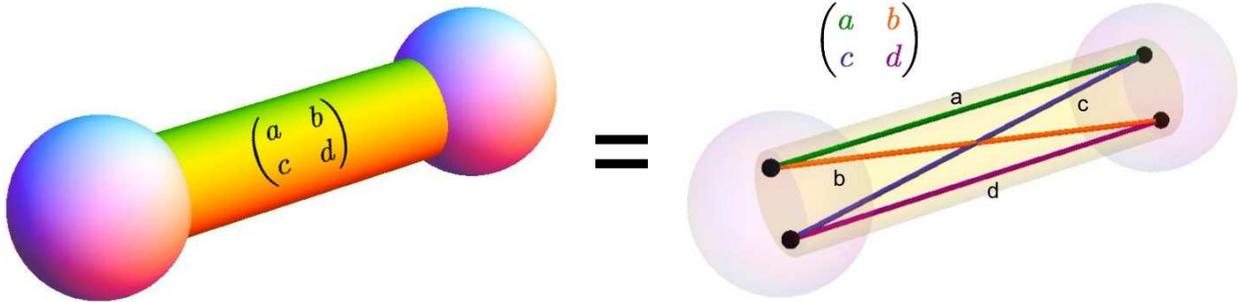}
\caption[Interpretation of a matrix-weighted edge]{Interpretation of a matrix-weighted edge.  A matrix-weighted edge connecting two macronodes specifies the graph connecting the underlying physical nodes.  In this example, the macronodes are connected by a $2 \times 2$ matrix-valued weight (left).  Each macronode itself therefore contains two physical nodes, and the real-valued weights connecting them are specified by entries in the matrix (right).
}
\label{fig:matrixweight}
\end{center}
\end{figure}

While real-valued weights cannot satisfy Eq.~\eqref{eq:orthograph} for a ring graph, \emph{matrix-valued weights} can, as shown below.  In such a case, Eq.~\eqref{eq:orthograph} becomes
\begin{align}
\label{eq:orthomatrixweights}
	\sum_l \mat A_{jl} \mat A_{lk} = \delta_{jk} \mat 1\;,
\end{align}
where the ``entries''~$\mat A_{jl}$ are themselves $m \times m$ matrices.  This means $\mat A$ is now an adjacency matrix on an $(mN)$-node graph.  On the other hand, treating the $m \times m$ blocks as single entries, $\mat A$ is the matrix-weighted adjacency matrix for what we call a \emph{supergraph}, which has \emph{macronodes} consisting of $m$~individual nodes each.  Figure~\ref{fig:matrixweight} illustrates the meaning of matrix-weighted edges between two macronodes.

\begin{figure}
\begin{center}
\includegraphics[width=\columnwidth]{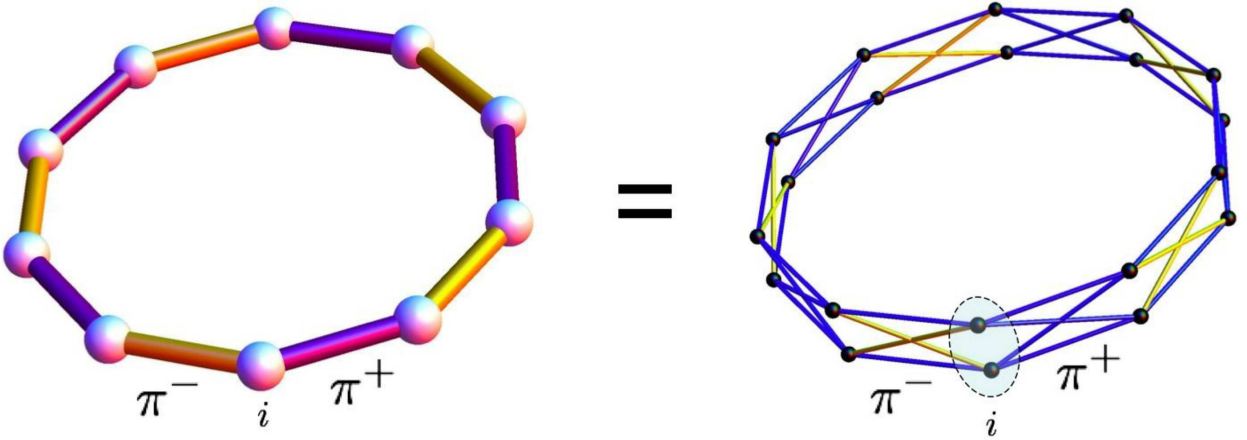}
\vspace{-.1in}
\caption[Matrix-valued weights and supergraphs]{Matrix-valued weights and supergraphs. The matrix-valued weights~$\pi^\pm$, defined in Eq.~\eqref{eq:Pidef}, connect the macronodes of a ``ring" supergraph (left).  The entries in $\pi^\pm$ specify the real-valued weights in the actual ``crown" graph (right) that connects the underlying physical nodes.  Measuring $q$ for each of the physical nodes in the top layer of the crown leaves the bottom layer in a uniformly weighted ring-graph CV cluster state.}
\label{fig:crown}
\end{center}
\vspace{-.15in}
\end{figure}

Promoting the ring graph to a supergraph with 2 nodes per macronode (i.e.,\ $m=2$), we can choose our weighting to consist of two orthogonal projectors over~$\reals^2$, labeled~$\pi^\pm$, with $\mat A_{j,j+1} = \pi^+$ for $j$~even and $\mat A_{j,j+1} = \pi^-$ for $j$~odd, where the total number of macronodes~$N$ is even, and index addition is modulo~$N$.  By reasons of symmetry and connectivity of the underlying graph, we choose $\pi^\pm$ to be
\begin{align}
\label{eq:Pidef}
	\pi^+ = \frac 1 2
	\begin{pmatrix}
		+ &	+ \\
		+ &	+
	\end{pmatrix}
	\qquad \text{and} \qquad
	\pi^- = \frac 1 2
	\begin{pmatrix}
		+ &	- \\
		- &	+
	\end{pmatrix}\;,
\end{align}
with $\pm$ standing for $\pm 1$.  With these weights, the \emph{supergraph} is still a ring, but the actual graph on physical nodes has a more complicated ``crown'' structure, as shown in Figure~\ref{fig:crown}.  Either by direct verification of Eq.~\eqref{eq:orthomatrixweights} or by noting that the geometric conditions~(1) and~(2) are satisfied, this weighting of the ring supergraph results in an orthogonal~$\mat A$.
Given a CV cluster state with this graph, one can perform local measurements on one of the two rings of physical nodes and measure it down to a CV cluster state~\cite{Menicucci2006} having a simple ring graph with equal weights, which is universal for single-mode operations using one-way QC methods~\cite{vanLoock2007a,Raussendorf2001}.  The only thing remaining is to show that the adjacency matrix for the crown can be put into Hankel or block-Hankel form by an appropriate numbering of the nodes.

It is easiest to begin at the supergraph level.  The ring is a circulant supergraph, which means it can be represented by a circulant adjacency matrix simply by numbering the nodes sequentially around the ring.  What we want, however, is the skew-circulant form, a special case of Hankel where the skew-diagonal bands ``wrap around'' at the edges of the matrix.  We can renumber the nodes to achieve this form as follows.  Starting with an arbitrary node, label it~1.  Move two nodes clockwise, and label that node~2.  Continue labeling every other node sequentially in a \emph{clockwise} fashion until you reach node $N/2$.  Then, pick one of the remaining nodes to label as $N/2+1$, move two nodes \emph{counterclockwise}, and label that node $N/2+2$, continuing in this fashion until all nodes are labeled.  The result is a skew-circulant adjacency matrix at the supergraph level, with matrix-valued weights given by Eq.~\eqref{eq:Pidef}.  Such a matrix is block-skew-circulant (and thus block-Hankel) at the physical node level:
\begin{align}
\label{eq:blockHankelring}
	\mat A \cong [\underbrace{0,\dotsc,0}_{N-3},\pi^+,0/\pi^-/\underbrace{0,\dotsc,0}_{N-3},\pi^+,0]\;,
\end{align}
where the zeroes are $2 \times 2$ blocks of all zeros, and $\cong$~indicates equality up to renumbering of nodes.

We could stop here, since we have an adjacency matrix that is block-Hankel with $2\times 2$ blocks.  The Hamiltonian~$\mathcal H(\mat A)$ from \eref{eq:Hamiltonian} can be implemented using a pump beam with three polarized frequencies and a PPKTP crystal~\cite{Pooser2005} using the method described in Section~\ref{S:OPOs}.  We will find it useful, however, to introduce a renumbering scheme that can be used to make the crown fully Hankel because a similar scheme will be used in the next section to reduce a block-Hankel matrix with $4\times 4$ blocks to one with $2\times 2$ blocks.  

Since $\mat A$~is an $N \times N$~matrix of $2 \times 2$~blocks, any index~$j$ can be thought of as an ordered pair $[m,c] := 2m + c = j$, with $m$ ranging from~$0$ to~$N-1$ and~$c$ being either~0 or~1. The permutation that reverses the order of the subindices, $[m,c] \to [c,m] := Nc + m = j'$, can be used to convert~$\mat A$ into fully Hankel form:
\begin{align}
\label{eq:Hankelring}
	\mat A \cong \frac 1 2 [\vec 0^{N-3},1,0,1,\vec 0^{N-3},1,0/{-1}/\vec 0^{N-3},1,0,1,\vec 0^{N-3},1,0]\;,
\end{align}
where $\vec 0^n$ represents a string of $n$~zeros.  Being fully Hankel, Eq.~\eqref{eq:Hankelring} can be used to directly read off the seven pump frequencies needed to implement this CV cluster state in a single OPO using an optical frequency comb of qumodes by the correspondence shown in Figure~\ref{F:Hankel}.  
Notice that since the only effect of increasing $N$ is to add more zeros to this Hankel vector---shifting, but not increasing in number, the required pump frequencies---the complexity of the pump is \emph{constant} with respect to~$N$, making this construction extremely scalable.

The CV cluster state described above therefore fulfills the four criteria desired in this section: it is \emph{bipartite}, it has an \emph{orthogonal} adjacency-matrix representation, that representation is also \emph{Hankel}, and it can be used to implement \emph{any single-mode operation}.  The first property satisfies the necessary condition for generation in a single OPO~\cite{Menicucci2007}. The second satisfies our desire for mathematical simplicity, allowing us to use the cluster-state graph~$\mat A$ without modification as the $\mathcal H$-graph from Eq.~\eqref{eq:Hamiltonian}.  The third guarantees simple experimental implementation~\cite{Zaidi2008}, with constant pump beam complexity.  The fourth says that single-mode measurements can be used to effect any single-mode unitary operation.  While not universal for multimode QC over continuous variables, the construction described here has introduced all the essential concepts for a construction that achieves this goal, described next.

\section{QC-Universal CV Cluster State}\label{S:torus}

The construction in the previous section gave a spectacularly efficient method of generating a CV cluster graph with one-dimensional topology (such as a line or ring), hence only suitable for single-mode operations. In universal one-way QC, however, operations involving more than one qubit/qumode require graph connections between such one-dimensional ``quantum wires.'' The natural candidate for a fully universal CV cluster-state graph is the square lattice---the original graph proposed for qubit cluster states~\cite{Raussendorf2001}---but, similar to the case made against the line graph, a square lattice does not admit an orthogonal adjacency matrix because it is not regular at the boundaries.  A logical response is then to impose toroidal boundary conditions, linking nodes on one side to those on the opposite side.  However, the  resulting toroidal square lattice suffers from the same problem as the simple ring: there exist pairs of nodes connected by exactly one 2-path, prohibiting an orthogonal representation.

\begin{figure}
\begin{center}
\includegraphics[width=.75\columnwidth]{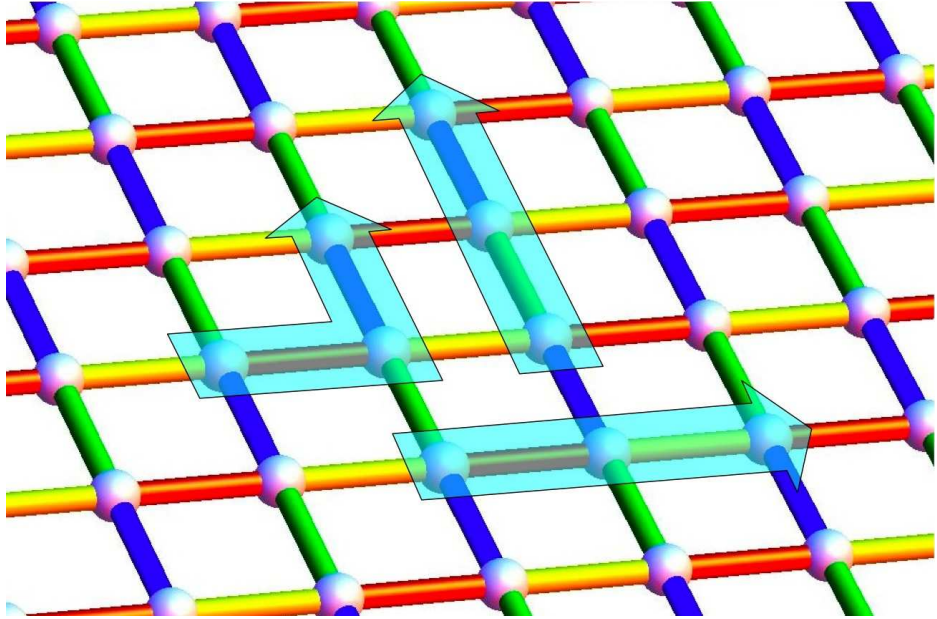}
\caption[Four-color solution to the geometric orthogonality conditions]{Four-color solution to the geometric orthogonality conditions.  Labeling the edges of a large lattice with four colors as shown, where each color represents one of four orthogonal projectors over~$\reals^4$, satisfies the geometric orthogonality requirements (see text).  From each node protrudes exactly one edge of each color, satisfying condition~(1), and any 2-path connecting distinct nodes (several shown) traverses edges with two different colors, which has exactly zero weight, satisfying condition~(2).}
\label{fig:latticeproj}
\end{center}
\end{figure}

We therefore take the same measures and promote the toroidal lattice to a supergraph.  Recall that orthogonality requires that $\mat A^2 = \mat 1$, equivalent to the aforementioned two geometrical conditions~(1) and~(2), which are local on the (super-)graph and thus are not affected by the overall topology.  In this case, the supergraph locally looks like a square lattice, which has degree four, and Figure~\ref{fig:latticeproj} illustrates that four mutually orthogonal projectors are sufficient to satisfy the geometric conditions.  Once again motivated by symmetry and connectivity of the underlying graph, we choose the weights to be $4 \times 4$ projectors constructed from the $2 \times 2$ projectors in Eq.~\eqref{eq:Pidef}:
\begin{align}
\label{eq:Pidefs}
	\Pi^0
	= \pi^+ \otimes \pi^+ 
	= \tfrac 1 4
	\left(\begin{smallmatrix}
		+ &		+ &		+ &		+ \\
		+ &		+ &		+ &		+ \\
		+ &		+ &		+ &		+ \\
		+ &		+ &		+ &		+
	\end{smallmatrix}\right)
	\;, \qquad
	\Pi^1
	= \pi^+ \otimes \pi^-
	= \tfrac 1 4
	\left(\begin{smallmatrix}
		+ &		- &		+ &		- \\
		- &		+ &		- &		+ \\
		+ &		- &		+ &		- \\
		- &		+ &		- &		+
	\end{smallmatrix}\right)
	\;, \nonumber \\
	\Pi^2
	= \pi^- \otimes \pi^+
	= \tfrac 1 4
	\left(\begin{smallmatrix}
		+ &		+ &		- &		- \\
		+ &		+ &		- &		- \\
		- &		- &		+ &		+ \\
		- &		- &		+ &		+
	\end{smallmatrix}\right)
	 \;, \qquad
	\Pi^3
	= \pi^- \otimes \pi^-
	= \tfrac 1 4
	\left(\begin{smallmatrix}
		+ &		- &		- &		+ \\
		- &		+ &		+ &		- \\
		- &		+ &		+ &		- \\
		+ &		- &		- &		+
	\end{smallmatrix}\right)
	\;.
\end{align}
We now possess an orthogonal supergraph, with 4-node macronodes, that looks locally like a square lattice but has toroidal topology.  The final task is to convert the resulting adjacency matrix into Hankel form.

Some freedom remains in the way that the lattice is rolled up into a torus that will turn out to make a Hankel form of the supergraph particularly easy to implement.  The freedom lies in the fact that we can {\em twist\/} the lattice as we roll it up: instead of connecting the last node of each line of the lattice to the first node in that same line, we can instead connect it to the first node of a different line---one that is $n$ lines away from the original one---generating a twist of $n$ lines in the torus.  Finding a Hankel form for the ring relied on the fact that the graph is circulant, along with a suitable node renumbering procedure, to generate a skew-circulant adjacency matrix.  While we could start with a toroidal lattice and try to find a twist that admits a Hankel form for it, we will instead explicitly construct a circulant graph and show that it is in fact a twisted toroidal lattice.  We will construct the graph from two different circulant ``threadings'' of $M^2$ nodes, corresponding to the two different dimensions of the lattice.
\begin{figure}
\begin{center}
\includegraphics[width=.6\columnwidth]{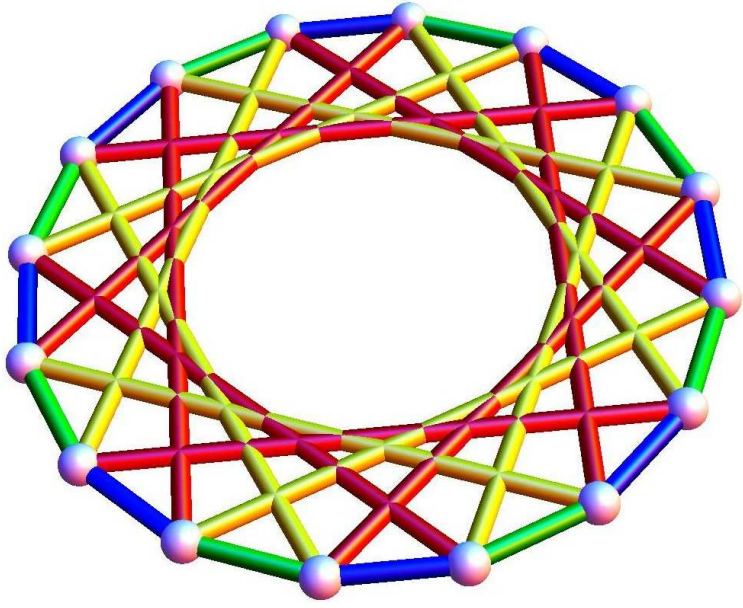}
\caption[Circulant embedding of a twisted toroidal lattice]{Circulant embedding of a twisted toroidal lattice.  The 16~white spheres each represent a four-node macronode, while the colored edges each represent a matrix-valued weight~$\Pi^j$ from Eq.~\eqref{eq:Pidefs}, with $\Pi^0$~red, $\Pi^1$~yellow, $\Pi^2$~blue, and $\Pi^3$~green.  The toroidal ``axes'' are identified with moving along the ``red-yellow direction'' and the ``blue-green direction.''  The circulant construction includes a one-unit twist in each toroidal dimension and generalizes easily to more nodes (see text).}
\label{fig:circulantlattice}
\end{center}
\end{figure}
\begin{figure}
\begin{center}
\includegraphics[width=1.00\columnwidth]{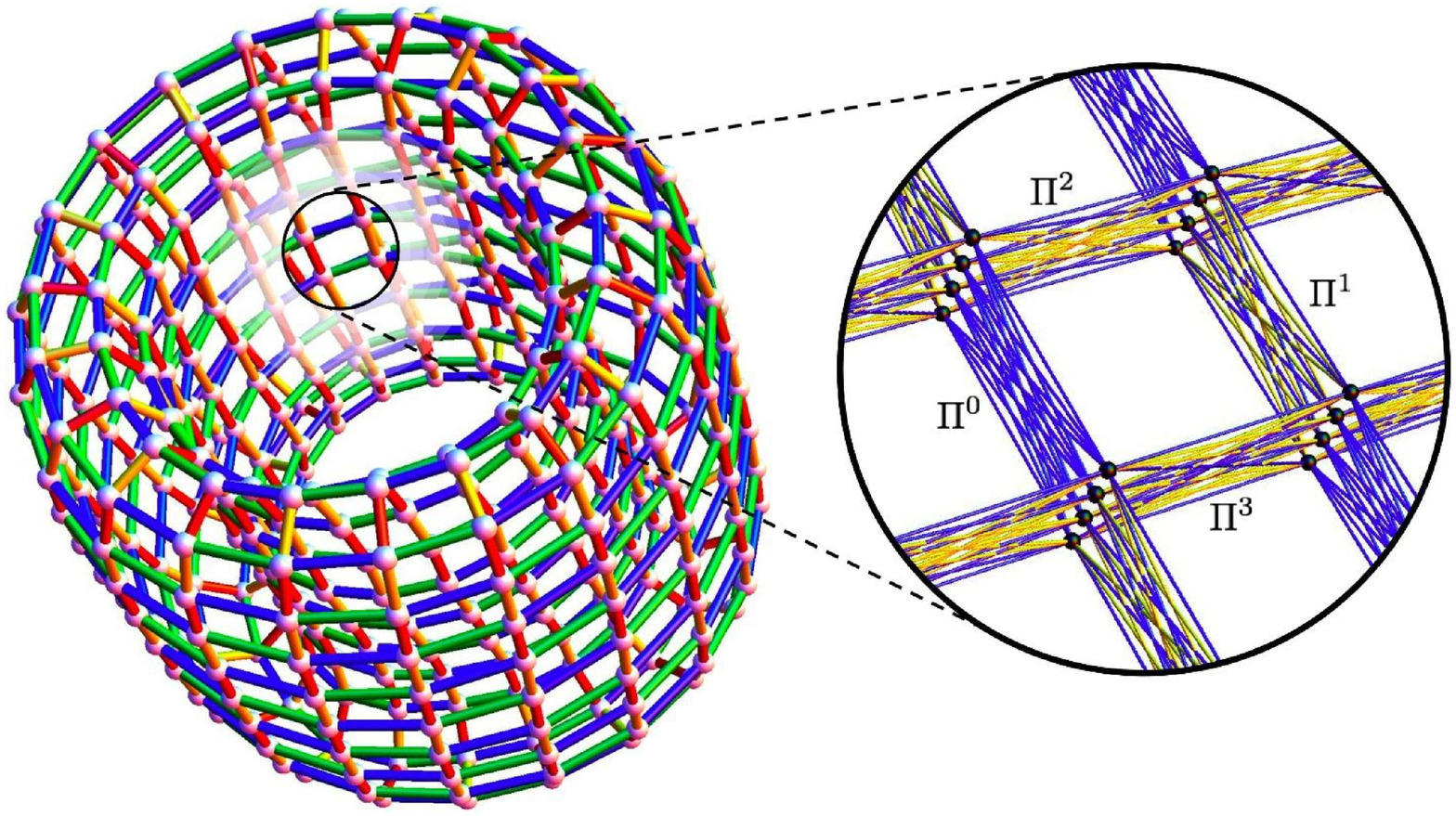}
\caption[Toroidal lattice supergraph and underlying graph structure]{Toroidal lattice supergraph and underlying graph structure.  Each of the $22^2$~macronodes in the supergraph (left) consists of four physical nodes, and each color corresponds to one of the four matrix-valued weights~$\Pi^j$ from Eq.~\protect\eqref{eq:Pidefs}.  Entries in $\Pi^j$ specify the real-valued weights connecting the underlying physical nodes (right).  Measuring $q$ on each physical node in three of the four ``layers'' leaves the remaining layer in a uniformly weighted QC-universal toroidal lattice cluster state.}
\label{fig:ttsl}
\end{center}
\end{figure}

The simplest circulant threading is just an $M^2$-node ring graph (we require $M$ to be even).  Alternating between two of the projectors from Eq.~\eqref{eq:Pidefs}, say $\Pi^2$ and $\Pi^3$, this is exactly the ring supergraph from the previous section (except now with 4-node macronodes).  We then apply our second ``threading'' by additionally connecting each node in the ring to the two nodes that are $M+1$ steps away, alternating with the remaining two projectors, in this case $\Pi^0$ and $\Pi^1$ (see Figure~\ref{fig:circulantlattice}).  Since $M+1$ and $M^2$ are relatively prime, this procedure creates another complete cycle through the ring, this time traversing every node by taking steps of size $M+1$.  The result is a circulant graph where each node now has four neighbors.  While tracing any path through the nodes, each step will move in one of the two threading ``directions,'' moving by $\pm 1$ or $\pm (M+1)$.  These correspond to moving horizontally or vertically, respectively, through the lattice.  The circulant nature of the graph guarantees toroidal boundary conditions.  Thus, we are left with a twisted toroidal square lattice supergraph, shown in Figure~\ref{fig:ttsl}.

\begin{figure}
\begin{center}
\includegraphics[width=.6\columnwidth]{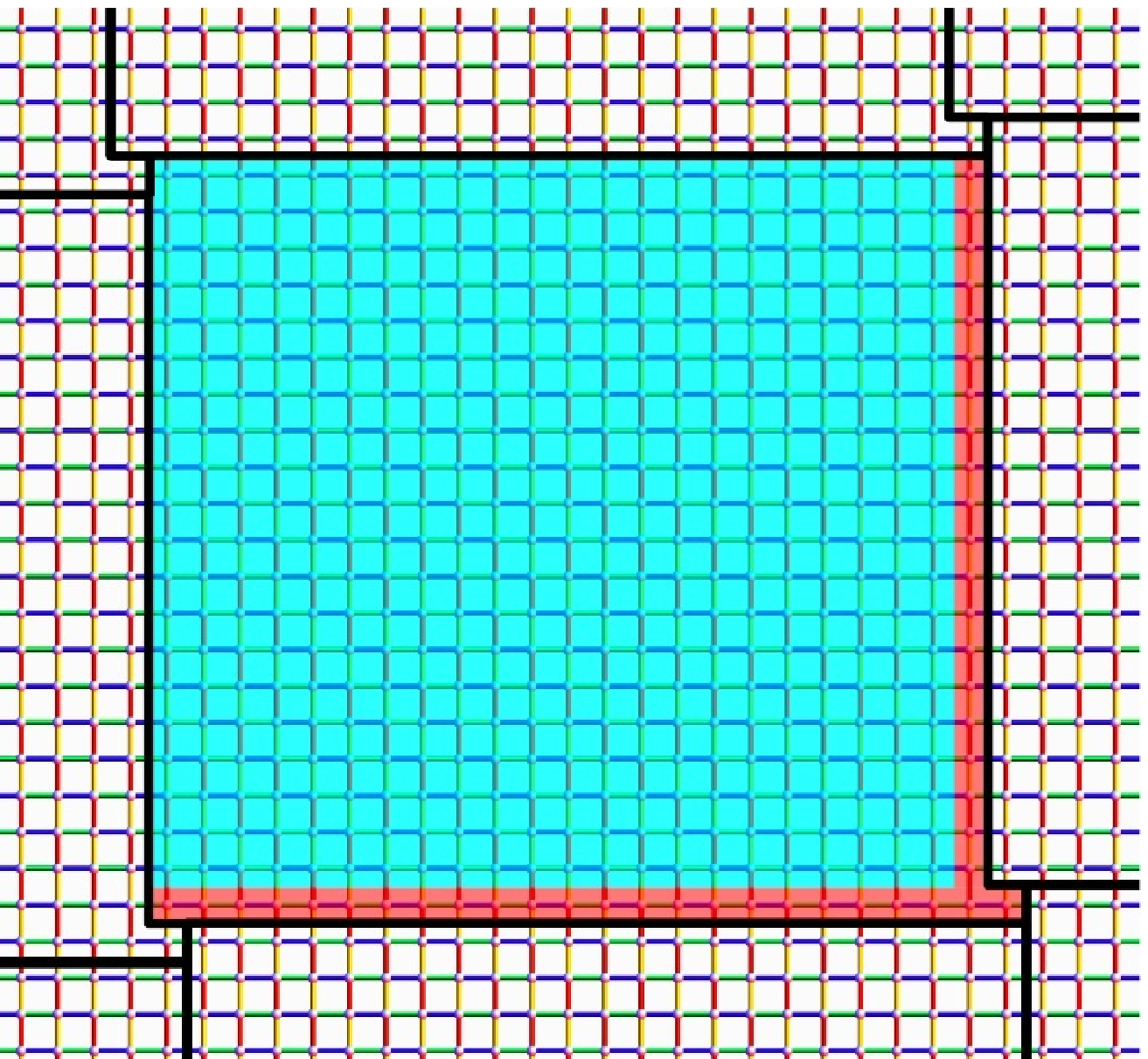}
\caption[Unrolling the torus]{Unrolling the torus.  The toroidal lattice in Figure~\ref{fig:ttsl} can be considered as a particular identification of macronodes on an infinite planar lattice that has the local structure of Figure~\ref{fig:latticeproj}.  The identification is specified by the (nearly) rectangular region delimited by black borders, tessellating the entire plane.  Macronodes at the same location with respect to their containing region are considered to be the same macronode; this enforces the toroidal boundary conditions.  Each region has a one-unit ``foot'' sticking out to account for the one-unit twist in each dimension (without the twist, the region would simply be square).  The region is~$(M+1) \times (M-1)+1$ macronodes in size (with the~$+1$ accounting for the ``foot''); the $23\times 21+1$~region shown corresponds to the $22^2$-macronode lattice in Figure~\ref{fig:ttsl}.  Measuring~$q$ on each node of the macronodes highlighted red ``unrolls'' the torus into an ordinary lattice of size~$M\times (M-2)$, highlighted blue.  The $M$-indexed family of such lattices is universal for QC; thus, the original toroidal lattice is, as well.}
\label{fig:unrolling}
\end{center}
\end{figure}

Since this graph is circulant, we can use the ring renumbering trick from the previous section (with $N \to  M^2$) to generate a skew-circulant adjacency matrix for this supergraph, which is block-skew-circulant (and thus block-Hankel) with $4 \times 4$ blocks at the level of the physical nodes:
\begin{align}
\label{eq:4x4skewcircTTSL}
	\mat A \cong [\vec 0^u,\Pi^1,\vec 0^v,\Pi^0,\vec 0^u,\Pi^3,0/\Pi^2/\vec 0^u,\Pi^1,\vec 0^v,\Pi^0,\vec 0^u,\Pi^3,0]\;,
\end{align}
where $u=(M-1)$ and $v=(M^2-2M-3)$.  Once again, the resulting matrix is block-Hankel instead of fully Hankel, this time with $4\times 4$ blocks.  For the ring, we used a renumbering of the nodes to convert the $2 \times 2$-block-Hankel matrix from Eq.~\eqref{eq:blockHankelring} into the fully Hankel matrix in Eq.~\eqref{eq:Hankelring}.   An equivalent renumbering can be done in this case to convert this $4 \times 4$-block matrix to one with $2 \times 2$ blocks, which {\em almost} preserves the Hankel structure (we'll show how to fix it below): simply treat the $4 \times 4$ blocks of $\mat A$ as $2 \times 2$-block matrices of $2 \times 2$ blocks, and apply the renumbering from the previous section (this time, with $N \to 2M^2$).  The resulting matrix is not completely block-Hankel, but it is close.  We can fix this by defining $\mat{A'}$ as in Eq.~\eqref{eq:4x4skewcircTTSL} but with the sign flipped in the next-to-last entry ($\Pi^3 \to -\Pi^3$).  While $\mat A$ is block-skew-circulant, $\mat A'$ is only block-Hankel (but still orthogonal):
\begin{align}
\label{eq:4x4blockHankelTTSL}
	\mat A' = [\vec 0^u,\Pi^1,\vec 0^v,\Pi^0,\vec 0^u,\Pi^3,0/\Pi^2/\vec 0^u,\Pi^1,\vec 0^v,\Pi^0,\vec 0^u,-\Pi^3,0]\;,
\end{align}
The important thing, though, is that if we apply the renumbering procedure to~$\mat A'$, the result is a $2 \times 2$-block-Hankel adjacency matrix:
\begin{multline}
\label{eq:2x2blockHankelTTSL}
	\mat{A'} \cong [\vec 0^s,\pi^-,\vec 0^t,\pi^+,\vec 0^s,\pi^+,0,\pi^-,\vec 0^s,\pi^-,\vec 0^t,\pi^+,\vec 0^s,\pi^-,0/{\pi^+}/ \\
	\vec 0^s,\pi^-,\vec 0^t,\pi^+,\vec 0^s,\pi^+,0,\pi^-,\vec 0^s,\pi^-,\vec 0^t,\pi^+,\vec 0^s,\pi^-,0]\;,
\end{multline}
where $s = (2M-1)$ and $t = (M^2-4M-3)$.

The $M$-indexed family of these cluster states is universal for CV one-way QC.  To see this, first cut open the toroidal lattice as shown in Figure~\ref{fig:unrolling}, to form an ordinary lattice.  Then, measure~$q$ on three physical nodes per macronode to reduce the supergraph to a uniformly-weighted graph with the same structure (see Figure~\ref{fig:ttsl}), which is known to be a universal graph~\cite{Menicucci2006}.  These cluster states are \emph{QC-universal}, \emph{bicolorable}, and \emph{orthogonal}---but still only \emph{block-Hankel}.  This is nevertheless sufficient for simple experimental implementation, using a method we now describe.

\section{Experimental implementation}\label{S:exp}

As already mentioned in Section~\ref{S:OPOs}, we propose to use optical polarization in order to implement the $2\times 2$ blocks~$\pi^\pm$. In Eq.~\eqref{eq:2x2blockHankelTTSL}, each $\pi^\pm$ block corresponds to a single pump frequency. The two diagonal elements of $\pi^\pm$ in \eref{eq:Pidef} correspond to interactions between two Y modes and between two Z modes. The two off-diagonal elements correspond to YZ and ZY interactions. 

Such couplings can be implemented in a nonlinear crystal such as PPKTP cut along the X axis. The point symmetry group of the crystal prescribes which elements of the nonlinear tensor are zero \cite{Yariv1985}. In the case of KTP, the nonzero elements are $d_{24}$, $d_{32}$, and $d_{33}$, which correspond to YZY/YYZ, ZYY, and ZZZ interactions respectively. (All other elements involve the X polarization and are therefore not accessible in an X-cut crystal.) The difference between $\pi^-$ and $\pi^+$ is thus a $180^\circ$ phase shift in the Y-polarized pump mode, and a narrowband pump polarized at $\pm 45^\circ$ in the YZ-plane implements a $\pi^{\pm}$ skew-diagonal band in~$\mat A$. Equation~\eqref{eq:2x2blockHankelTTSL} therefore translates into a single OPO pumped by exactly 15~frequencies, which must be phase-locked.  

The experimental demonstration of all three interactions simultaneously was performed in a PPKTP crystal with a $45.65\ \mu$m period, at a signal wavelength of 1490 nm \cite{Pooser2005}. The crystal's spatial modulation of its nonlinear tensor yielded quasiphasematching of the interactions  using different harmonics of the same poling period, a remarkable triple coincidence. However, simultaneity at any point in the crystal is not required. It suffices to have all interactions occurring every roundtrip in the cavity, since the fields undergo many of these before being coupled out, which leads to integration of the quantum fluctuations over the OPO cavity's storage time and to the well-known squeezing spectrum \cite{Walls1994}. (This does yield a finite bandwidth for quantum processing and can be circumvented by the use of short-cavity OPOs in order to increase the cold cavity linewidth.) Hence, a suitable nonlinear material can be straightforwardly designed by quasiphasematching each interaction independently, with its own poling period in separate regions of the crystal. This will allow us to use the first (most efficient) modulation harmonic for each interaction, as well as to tailor the respective lengths of the poled regions so as to compensate for the different values of the aforementioned nonlinear coefficients, and achieve the equal theoretical coupling strengths used in this paper. Note that unequal interaction strengths can be compensated by varying the relative pump power between the two polarizations, thereby providing a fine tuning mechanism for optimizing the entanglement generation. These design improvements to the nonlinear crystal are in progress.

Also, interactions with OFC modes outside the desired subset must also be strictly suppressed in order to realize the exact desired graph state. This can be done by cavity mirror design and/or by quasiphasematching bandwidth design~\cite{Fejer1992} and does not present a fundamental issue. While the pump spectrum is relatively complex---requiring 15~frequencies---that number is \emph{constant} with respect to the lattice size, making this construction extremely scalable.

Note that, in principle, additional physical parameters like wave vector direction or transverse mode structure could be used to directly implement block-Hankel $\mat A$'s with larger blocks (e.g.,~$4\times 4$).  This would reduce pump complexity but require a more sophisticated OPO, in which the nonlinear couplings between these additional quantum labels should be precisely defined. Other avenues, such as implementation of nonlinear couplings in slow light media, are conceivable. Of particular interest would be the use of nonlinear waveguides, seeded by the OFC from a modelocked laser. The interest here is to obtain nonlinear efficiency via the intensity enhancement from optical  confinement, while doing away with the resonant buildup of intracavity losses in an OPO. Recently, an improved, though still modest, amount of -2.2 dB squeezing was obtained in a PPKTP waveguide~\cite{Pysher2008}.

Finally, we examine the experimental requirements for scaling the CV cluster state with $N$~vertices. As we have shown, the pump spectrum is independent of $N$~and the OFC has constant and large size. The scaling requirements are twofold. 

First, the number of pump photons must increase with the number of entangled qumode pairs, or graph edges. The toroidal square lattice is a sparse graph, with $\cO(N)$~edges for $N$~vertices, in lieu of the $N(N-1)/2$~edges of a complete graph.  Increasing the number of vertices therefore yields a \emph{linear} increase of the number of edges, i.e.\ of the entangled qumode pairs. Thus the number of pump photons scales linearly with the number of entangled modes.  Since a two-mode squeezing experiment in a material like KTP can utilize as low as a few milliwatts of pump power and since KTP can withstand up to a few watts of pump power (continuous-wave, focused), this allows scaling by 3 orders of magnitude. This figure is only given as an example of what the possibilities are, without any optimization, in a typical nonlinear material with a fairly high (but not the highest) optical damage threshold.

Second, the bandwidth of the nonlinear coupling must encompass all entangled qumodes, that is, as the graph grows, the skew diagonals in $\mat A$ grow, again linearly with $N$.  The phasematching bandwidth is inversely proportional to the length of the nonlinear crystal (over which the nonlinear effect is integrated) hence gaining bandwidth by shortening the crystal would yield a reduction of the coupling strength and an optimum must be found. Examining again---and conservatively---typical figures for existing experiments gives a phasematching bandwidth of 100 GHz to 1 THz to be compared to an OPO FSR of 100 MHz ($1.5$ m long cavity) gives of the order of $10^3$--$10^4$ modes. It is therefore clear from the present estimations that our approach already has a significant realistic scaling potential, in line with the theoretical promise.

\section{Measurements}\label{S:measure}

Up to this point we have focussed entirely on creating the universal resource for one-way quantum computation, a CV cluster state.  Of course, any useful implementation also needs a viable scheme for performing local adaptive measurements.  In the original paper on CV one-way quantum computing~\cite{Menicucci2006}, the authors show that universality in an optical measurement-based scheme can be achieved with just two types of measurement: (1)~homodyne detection and (2)~a non-Gaussian projective measurement such as photon counting.  Homodyne detection is easily performed at close to unit efficiency, and the best photon number-resolving detectors currently operate around 95\% efficiency~\cite{Lita2008}, with the technology continually improving.  As a further simplification, recent results show that all photon-counting measurements can be performed first, transforming the cluster state into a non-Gaussian resource state~\cite{Gu2009}.  This is useful because all remaining measurements are homodyne detections, which are considerably simpler.  These measurements are no longer parallelizable~\cite{Menicucci2006,Raussendorf2002} because they are used to gate-teleport non-Gaussian operations onto the encoded quantum information in the cluster.

The unique aspect of our scheme as compared to other optical implementations using CV cluster states is that our method packs all qumodes into a single beam.  Thus, separately addressing the polarization- and frequency-labeled qumodes is an obstacle not present in implementations relying on spatially separated beams~\cite{vanLoock2007,Yonezawa2008,Yukawa2008,Su2007}.  While separating orthogonal polarizations is straightforward, doing so with neighboring cavity modes may seem difficult at first glance since an ordinary prism would not have enough resolving power to discriminate optical waves separated by radio frequencies. To get around this problem, one can take advantage of new tools for optical frequency combs such as the virtually-imaged phased array (VIPA) disperser, which is an interference filter that provides angular separation at the needed resolution and is compatible with lossless quantum optical designs~\cite{Shirasaki1996,Xiao2004,Diddams2007}.  Polarization optics and VIPAs can therefore be used to separately address each qumode with individual detectors (whose number necessarily scales linearly with the size of the quantum register in any implementation).

\section{Error Correction and Fault Tolerance}\label{S:qecft}

The main difficulty with any quantum computing scheme is preventing errors and decoherence. The theory of quantum error correction and fault tolerance~\cite{Nielsen2000} was developed for exactly these reasons.  While most of the existing literature deals with the case of qubit systems, or more generally, finite dimensional systems, our scheme must address the additional difficulties of working with continuous variables.  Fortunately, error correcting codes have been developed in the CV setting~\cite{Braunstein1998a, Gottesman2001, Barnes2004, Glancy2006, Niset2008, Beny2008}, usually by mapping the CV system into an effective qubit space~\cite{Gottesman2001, Ralph2003}.  In some cases, this can be used to establish a fault-tolerance threshold~\cite{Lund2008}.  Lower-dimensional encodings may be the only option for fault tolerance using CV cluster states, given the analogy with classical analog computation, where no error threshold exists.

The finite-squeezing approximation is a special case of these more general considerations of error correction and fault tolerance.  Certainly, more squeezing is preferable to less, but the amount required for any particular QC task remains an open question.  Consequently, it's unclear how the nonlinear coupling strength~$\kappa$ in Eq.~\eqref{eq:Hamiltonian} will need to scale with~$N$ for any particular QC application.  While finite squeezing errors can be mitigated in medium-sized proof-of-principle experiments with CV cluster states~\cite{Menicucci2006}, the task of rigorously establishing a fault-tolerance threshold for CV one-way QC is an important open problem.  We hope to spur further investigations along these lines. 

\section{Conclusion}\label{S:conclusion}

Quantum computation has come a long way in the last 15 years.  There are now multiple schemes---both experimental and theoretical---for processing quantum information, all of which show promise in different areas.  The particular scheme proposed here combines the power of cluster-state quantum computation, which harnesses entanglement to substitute for coherent unitary evolution, with the unorthodox model of quantum computation using continuous variables.  Experimentally, we propose an efficient method of generating very large continuous-variable cluster states using an exceptionally compact experimental setup: a single optical parametric oscillator does all the work, entangling all the desired modes at once.  One surprising feature of this scheme is that the complexity of the required pump spectrum is constant even as the cluster state is scaled up to include more modes.  Much of the experimental technology required already exists, and implementation is underway at the University of Virginia.

The entangled states produced by this method will also be objects of interest in the study of entanglement at mesoscopic scales.  Further simplification of the method may be possible using additional degrees of freedom, such as spatial modes.  This method of entangling---in one fell swoop---a large optical frequency comb into a continuous-variable cluster state is the first ``top-down'' approach proposed for one-way quantum computation using optical encodings of information in either discrete or continuous variables.  While open questions remain about the effects of finite squeezing on scalability for particular quantum computing tasks, the unprecedented scalability of this method is exciting and motivates further theoretical and experimental research.


\acknowledgments

The authors thank Henry Haselgrove, Earl Campbell, and Akimasa Miyake for discussions.  STF~was supported by the Perimeter Institute for Theoretical Physics.  OP~was supported by National Science Foundation grants No.\ PHY-0555522 and No.\ CCF-0622100.  NCM~was supported by the National Science Foundation and by Perimeter Institute.  NCM and OP are grateful to Perimeter Institute for its hospitality during productive and stimulating visits.  Research at Perimeter Institute is supported by the Government of Canada through Industry Canada and by the Province of Ontario through the Ministry of Research \& Innovation.


\bibliographystyle{bibstyleNCM}
\bibliography{longtorus}


\end{document}